\documentstyle[preprint,aps]{revtex}
\tightenlines
\begin{document}

\draft

\title{Interlayer Conductivity in the Superconductor
$Tl_{2}Ba_{2}CuO_{6+\delta}$: Energetics and Energy Scales}

\author{A.S.~Katz,$^{1}$ S.I.~Woods,$^{1}$ E.J.~Singley,$^{1}$ 
T.W.~Li,$^{2}$ \cite{MX} M.~Xu,$^{3}$ \cite{MX} D.G.~Hinks,$^{2}$ 
R.C.~Dynes,$^{1}$ and D.N.~Basov$^{1}$} 
 \address{$^{1}$Department of Physics, University of California, San Diego; 
La Jolla, CA  92093-0319} 
 \address{$^{2}$Department of Materials Science, Argonne National 
Laboratory, Argonne, IL  60439} 
 \address{$^{3}$James Franck Institute, University of Chicago, Chicago, IL  
60637} 

\maketitle

\begin{abstract}
We report on infrared studies of the $c$-axis electrodynamics of  
$Tl_{2}Ba_{2}CuO_{6+\delta}$ crystals.  A sum rule analysis reveals 
spectral weight shifts that can be interpreted as a kinetic energy change 
at the superconducting transition. In optimally doped crystals, showing an 
incoherent normal state response, the kinetic energy is lowered at $T<T_c$, 
but no significant change is found in the over-doped samples, which have more 
coherent conductivity at $T>T_c$. 
 \end{abstract} 


Despite extensive experimental effort in high-$T_c$ superconductivity over 
the last decade very little is known about the microscopic roots of the 
condensation energy $E_c$ in this class of materials. In conventional 
superconducting metals $E_c$ is due to the {\it reduction} of the potential 
energy, which overwhelms the {\it increase} of the electronic kinetic 
energy\cite{tinkham}.  Several models of high-$T_c$ superconductivity 
propose entirely different energetics of $E_{c}$ with  the superconducting 
state being driven  by changes of the Coulomb energy\cite{legget98}, 
exchange energy\cite{demler,white} or  kinetic 
energy\cite{hirsch87,anderson95,hirsch92,chakravarty98,emery99}. Existing 
experimental results are inconclusive\cite{moler98,tsvetkov,Basov99,demler}. 

A variety of models enable inference of the electronic kinetic energy  
from a sum rule analysis of the complex optical conductivity 
$\sigma(\omega)=\sigma_1(\omega)+i\sigma_2(\omega)$\cite{maldague}.  A 
partial sum rule  relates the integral of $\sigma_1(\omega)$ over a limited 
energy range $W$ to the kinetic energy $-\alpha$: 
 \begin{equation}
 \int_0^Wd\omega \, \sigma_1(\omega) = -\alpha. \label{KEeqn}
 \end{equation}
 For the case of a superconductor Eq.\ (\ref{KEeqn}) can be rephrased in 
the form of a modified Ferrel-Glover-Tinkham (FGT) sum rule to 
yield the kinetic energy sum rule\cite{hirsch92,chakravarty98,kim98}: 
 \begin{equation}
\rho_{s} = [N_{n} - N_{s}] + [\alpha_{n} - \alpha_{s}]  \label{FGT}
 \end{equation} 
In  Eq.\ (\ref{FGT}), $\rho_s$ is the 
superfluid density that quantifies the spectral weight under 
the superconducting $\delta$-function\cite{rhodef}. Integrals 
$N_n(\omega)=\int_{0^+}^{\omega} d\omega'\sigma_1(\omega', T>T_c)$ and 
$N_s(\omega)=\int_{0^+}^{\omega} d\omega'\sigma_1(\omega', T\ll T_c)$ are 
proportional to the number of carriers participating in absorption above 
and below $T_c$.  The magnitude of $\rho_s=4\pi\omega\sigma_2(\omega 
\rightarrow 0)$ and  $[N_n-N_s]$ are obtained independently from the 
optical constants\cite{Basov99}. Typically, the $[N_n-N_s]$ integrals  
converge at energies below 10-15 $k_{B}T_c$\cite{ab}.  A comparison of 
the magnitudes of $\rho_s$ and $[N_n-N_s]$ then provides a basis for an 
experimental probe of the kinetic energy change $\Delta \alpha = 
[\alpha_{n} - \alpha_{s}]$. Physically, the non-vanishing magnitude of  
$\Delta\alpha$ implies that changes  of the low-energy spectral weight in 
Eq.~(1)  are compensated by  readjustment of interband transitions at 
$\omega>W$  so that  the global sum rule $\int_0^\infty d\omega \, 
\sigma_1(\omega) = \frac{\pi n e^{2}}{2 m_{e}}$ is satisfied. 

Recently, we found a $\rho_s>[N_n-N_s]$ inequality in the interlayer 
conductivity of a variety of cuprates indicating that the superconducting 
condensate is collected from an interband-scale energy range\cite{Basov99}. 
This result can be interpreted in terms of a reduction of the kinetic 
energy below $T_c$.  In this paper we explore the connection between the 
development of coherence in the interlayer conductivity and the anomalous 
behavior of the c-axis superconducting condensate in 
$Tl_{2}Ba_{2}CuO_{6+\delta}$ (Tl-2201).  We found that the energy scale 
associated with superconductivity extended to the interband region when 
the normal state conductivity across the layers was nearly blocked. Once 
the interlayer transport above $T_c$ became more coherent, the FGT sum rule 
approached exhaustion at energies below $\sim$5~$k_{B}T_c$ suggesting that any 
kinetic energy change was very small or nonexistent. 

$Tl_{2}Ba_{2}CuO_{6+\delta}$  is an ideal material for studying the 
 over-doped regime of the phase diagram in which a variety of cuprates 
reveal a crossover to a more coherent response. Highly over-doped crystals of 
Tl-2201 have nearly identical chemical composition to the optimally doped 
phase since a less than 2\% change in the oxygen content is required to 
suppress $T_c$ from the maximum value of $T_{c} \simeq 90$~K down to less 
than 4 K\cite{kubo91}.  Crystal preparation is described 
elsewhere\cite{moler98}. Typical crystal dimensions were nominally
 0.8 mm x 0.8 mm x 0.04 mm.  Mosaics of several specimens with similar 
$T_{c}$ and $\Delta T_{c}$ (determined from magnetization measurements) 
were prepared and polished along a face parallel to the {\it c}-axis.  The 
polished surfaces were smooth, black, and shiny.  Infrared reflectance, 
$R_{c}(\omega)$, was measured with $E \parallel c$ polarized light in the 
frequency range from 16 to 15,000 cm$^{-1}$. Spectra taken at different 
temperatures had a relative experimental uncertainty of less than 0.5\%.  

In Fig.\ \ref{fig1}, we show the raw reflectance results for the optimally 
and over-doped crystals.  The $E \parallel c$ reflectance of both samples 
over the range of 80 to 2000 cm$^{-1}$ (10 to 250 meV) resembled an ionic 
insulator with a low reflectivity punctuated by phonon peaks. Both samples 
showed a plasma edge below $T_{c}$.  Classical electrodynamics describes 
the development of this plasma edge since superconducting currents flow 
along all crystallographic directions\cite{shutzmann97}. In the optimally 
doped sample ($T_{c} = 81$ K, $\Delta T_{c} \simeq $ 8 to 10 K) the plasma 
edge grew out of an apparently insulating normal state spectrum.   In the 
over-doped sample ($T_{c} = 32$ K, $\Delta T_{c} \simeq 5$ K) we found a 
``metallic'' up-turn in $R_c(\omega)$ measured above $T_{c}$, consistent 
with more coherent delocalized behavior.  Below $T_{c}$, the reflectance 
continued to rise, and the plasma edge developed a characteristic minimum 
at a frequency position of $\omega = 49 \text{ cm}^{-1}$.  The minimum in 
reflectance was shifted to higher frequencies compared to $\omega = 37 
\text{ cm}^{-1}$ in the optimally doped case and was less sharply defined. 
 
The raw $E \parallel c$ reflectance data was transformed using the 
 Kramers-Kronig relations to determine the complex conductivity. 
Extrapolations of the reflectance data to low and high frequencies, 
required for the Kramers-Kronig integrals, did not strongly affect the 
results in the frequency range where measured data exists. The real part 
of the conductivity spectra were dominated by strong phonon peaks (Fig.~2). 
Three of the modes at 85, 151, 360 cm$^{-1}$ were unaffected by doping; the 
highest frequency mode softened from 602 cm$^{-1}$ in the optimally doped 
crystal down to 595 cm$^{-1}$ in the over-doped sample.  The electronic 
background of the conductivity of the optimally doped sample was nearly 
flat and featureless for temperatures above and below $T_{c}$.  The 
response of the over-doped crystal was different. The conductivity below 80 
cm$^{-1}$ showed a Drude-like behavior steadily rising out of a $\sim$5 
$(\Omega \text{ cm})^{-1}$ background.  Below $T_{c}$, this ``metallic'' 
response diminished as temperature was lowered.  By T = 5 K, it 
effectively vanished, and the conductivity (ignoring 
phonon peaks) was nearly frequency independent.  Just above $T_{c}$, the dc 
conductivity of the over-doped sample at T = 35 K (obtained from the 
extrapolation of $\sigma_1(\omega)$ to $\omega=0$) was $\sim$15 
$(\Omega \text{ cm})^{-1}$.  This is roughly a factor of 3 higher than the 
value obtained for optimally doped Tl-2201 crystals.  It has been 
established that increasing oxygen content causes the {\it c}-axis to 
contract\cite{contract}. This closer spacing of the $CuO_{2}$ planes favors 
enhanced interlayer coupling and likely leads to the more coherent behavior 
of the conductivity. It is important to emphasize, however, that the 
$c$-axis conductivity is still two orders of magnitude smaller than the 
in-plane dc conductivity, and it is 4 to 6 orders of magnitude smaller than 
the conductivity of conventional metals. 

Using the complex conductivity spectra above and below $T_{c}$, we can 
identify the spectral origins of the superfluid condensate.  In all of our 
measurements, the conductivity at $T < T_{c}$ was suppressed compared to 
spectra taken at $T \simeq T_{c}$, but we saw no evidence of a classical 
superconducting gap $\Delta$ for $T \ll T_{c}$.  The absolute value of 
$\sigma_{1}(\omega)$ persisted  well above the noise level down to the lowest 
measured frequencies, behavior consistent with gaplessness.  To quantify 
the transfer of spectral weight below $T_c$, we plotted the ratio 
$[N_{n}(\omega) - N_{s}(\omega)] / \rho_{s}$ as a function of $\omega$ 
in Fig.~3\cite{Basov99}. The superfluid density $\rho_s$ was determined from 
the extrapolation of $\omega \sigma_{2}(\omega,T)$ to $\omega = 0$ for 
$T \ll T_{c}$, a procedure that does not require model-dependent assumptions 
\cite{values}.  The frequency dependence of
 $[N_{n}(\omega) - N_{s}(\omega)] / \rho_{s}$ unfolds 
the sum rule integrals.  In the optimally doped sample, the ratio rose 
slowly and saturated near a value of about 0.6 at $\omega\simeq 30 - 40$ 
meV.  The energy interval in Fig.\ \ref{fig3} corresponds to $\sim$22 
$k_{B}T_c$, 
 more than sufficient for the conventional FGT sum rule. The data for the 
optimally doped sample implied that a significant portion of $\rho_{s}$ 
was accumulated from energies above 0.15 eV. In contrast, in over-doped 
 Tl-2201 the dominant fraction of the superfluid density was collected  from 
the far-infrared, low energy region.  Within our experimental uncertainty, 
at least 80 to 90\% of $\rho_s$  was accumulated from energies as small as 
4 - 5 $k_{B}T_c$\cite{phononsnote}. Comparison of the data plotted in Fig.\ 
\ref{fig2} and in Fig.\ \ref{fig3} suggests that the behavior of the 
conductivity above $T_c$ determines the energy scales from which the 
superconducting condensate is collected.  In the case of the over-doped 
crystal there is no need to extend integration of the conductivity to 
 mid-infrared and near-infrared energies in order to account for the 
magnitude of $\rho_s$ since the required spectral weight is readily 
available at low energies.   

If Eq.~(2) is chosen as the basis for data interpretation, then the 
$\rho_s>[N_n-N_s]$ inequality observed in the optimally doped crystal  
implies a reduction  of the electronic kinetic energy below $T_c$. The 
reduction of the kinetic energy can be understood by noting that the 
charge carriers are nearly confined to the CuO$_2$ planes above $T_c$ 
whereas in the superconducting state {\it paired}  charges move more 
easily between the layers. On the contrary, the over-doped sample, 
which revealed a more coherent $c$-axis response (Drude-like upturn of 
$\sigma_1(\omega)$ and enhanced dc conductivity), showed 
$\Delta  \alpha\simeq 0$  within our error bars. The 
data suggested that a kinetic energy change is observed only if 
deconfinement of the charge carriers occurs by virtue of pair tunneling 
below $T_c$.  If the charge carriers were already delocalized in the normal 
state (even with very low plasma frequencies), then the magnitude of 
$\Delta \alpha$ is vanishingly small. The same trend was observed in the 
$c$-axis response       of the YBa$_2$Cu$_3$O$_x$ (YBCO) family. The 
discrepancy between $[N_{n}(\omega)-N_{s}(\omega)]$ and 
$\rho_{s}$, suggesting a lowering of the kinetic energy,  
was most extreme in YBa$_2$Cu$_3$O$_{6.53}$. As the oxygen content was 
raised, the excess spectral weight appeared in the  lower energy part of 
$\sigma_1(\omega, T>T_c)$ spectra. The $[N_{n}(\omega)-
N_{s}(\omega)]/\rho_{s}$ ratio with integration limited up to 10-15 
$k_{B}T_c$ was near unity for $YBa_{2}Cu_{3}O_{6.85}$ and higher doping 
levels indicating that $\Delta\alpha$ became vanishingly 
small\cite{ybcounpub}. 
  
Further analysis of the over-doped Tl-2201 crystal showed a lower energy 
scale associated with superconductivity. It has been experimentally 
observed in a wide variety of cuprates that the magnitude of the 
 {\it c}-axis penetration depth $\lambda_{c} = c/\sqrt{\rho_{s}}$ is 
related to the dc conductivity $\sigma_{dc}$ at $T \simeq T_{c}$ by the 
relationship 
 \begin{equation} \lambda_{c}^{-2} = \frac{1}{\hbar 
c^{2}} \Omega_{S}\, \sigma_{dc}(T \simeq T_{c}), 
\label{LamEq}
 \end{equation}
 where $\Omega_{S}$ is an energy scale associated with 
superconductivity\cite{basov94,uchida96,kirtley98,chakravarty98}.    
Eq.\ (\ref{LamEq}) may be obtained either by modeling a dirty limit bulk 
superconductor or by treating the anisotropic cuprates as a weakly coupled 
stack of intrinsic Josephson junctions.  Those models find $\Omega_{S} = 
4\pi^{2}\Delta$. In our experiments, $\lambda_{c}$ remained nearly 
unchanged \cite{values} whereas the $c$-axis dc conductivity increased by a 
factor of $\sim$3 in the over-doped crystals. 
 If Eq.\ (\ref{LamEq}) holds true, then $\Omega_{S}$ must be reduced in the 
over-doped regime in order to keep the penetration depth constant. 
Supporting evidence that $\Omega_{S}$ is indeed reduced for over-doped Tl-
2201 comes from Raman scattering measurements on optimally and over-doped 
Tl-2201 which found that a spectral feature attributed to the superconducting 
energy gap $2\Delta$ shifted down in frequency from $\sim$350~cm$^{-1}$ 
in optimally doped Tl-2201 to $\sim$105~cm$^{-1}$ in an over-doped sample 
as $T_{c}$ fell from 78 K to 37 K\cite{kendziora96,kang96}. {\it c}-Axis
polarized Raman scattering in Bi$_{2}$Sr$_{2}$CaCu$_{2}$O$_{8+\delta}$ (Bi-2212)
single crystals also showed an energy scale that diminished in over-doped
samples\cite{liu99}.
 The lower energy 
scale of the superfluid condensation seen in Fig.~3 is consistent with 
a reduction in $\Omega_{S}$\cite{footnote2}. 
                                                          
Several important distinctions of over-doped phases from the 
optimally-doped counterparts have been observed in different 
families of the cuprates.  Raman spectroscopy showed both in Tl-2201 and 
Bi-2212 single crystals highly 
anisotropic energy gaps at optimal doping but revealed a crossover to 
isotropic behavior in  over-doped samples\cite{kendziora96}. Consistent 
with this latter result, angular resolved photoemission (ARPES) on Bi-2212 
found that the Fermi surface nodes vanished in the over-doped 
regime\cite{kelley96}. Thus, both Raman and ARPES studies suggested a trend 
towards the development of a more isotropic superconducting state in 
over-doped crystals. We observed a more conventional superfluid response in 
over-doped Tl-2201 compared to the optimally-doped sample. Phase sensitive 
measurements of the order parameter in over-doped materials are needed to 
determine if the above observations are associated with the development of 
an $s$-component of the order parameter.   An {\it s}-wave component in the 
order parameter in the over-doped regime is possible within the 
 stripe-based models\cite{emery99}.  
 
 Based on the existing experimental results we were unable to correlate 
 changes in the kinetic energy with the critical temperature of the studied 
 superconductors. In the Tl-2201 series, $\Delta\alpha$ is largest in the 
optimally-doped crystals. The YBCO data, however, revealed large 
changes of $\alpha$ in the underdoped regime whereas in the optimally 
doped samples the effect was negligible\cite{Basov99}. The common parameter 
determining the kinetic energy change for both  series of crystals appears 
to be the dc conductivity. It remains to be seen if the absolute value of 
$\Delta\alpha$ is sufficient to account for the condensation energy 
determined from  specific heat 
measurements\cite{moler98,tsvetkov,chakravarty98}.

In conclusion, we employed infrared spectroscopy to examine the energy 
scales of superconducting state in Tl-2201.  We found that the 
superfluid spectral weight was accumulated from lower energies when the 
 $c$-axis conductivity showed a more coherent, delocalized response. Using 
the modified FGT sum rule we could extract kinetic energy changes and 
correlate this change to the normal state $c$-axis conductivity.

We thank D.A. Kossenko for technical assistance.  Work at UCSD supported 
by NSF Grant, DMR-9875980, AFOSR Grant F4962-098-0264, DOE Contract 
DE-AC02-98CH10886, the Sloan Foundation, and the Research Corporation.  
Work at Argonne supported by NSF Grant DMR 91-20000 and by DOE Contract 
W-31-109-ENG-38.

\begin{figure}
\caption{Reflectance of Tl-2201 measured with $E \parallel c$ polarization 
of incident radiation.  Left panel: The optimally doped sample develops a 
superconducting plasma edge out of an apparently insulating spectrum for $T 
< T_{c}$.  Right panel: At the lowest energies, the over-doped sample shows 
continuously increasing reflectance above and below $T_{c}$.} \label{fig1} 
\end{figure}

\begin{figure}
\caption{Real part of the complex conductivity of Tl-2201 calculated from 
the reflectivity shown in Fig.~1.  Left panel: The conductivity 
of the optimally doped sample is punctuated by phonon peaks but is 
otherwise nearly flat and featureless.  Right panel: The over-doped sample 
shows the same phonon peaks but develops a coherent, Drude-like peak for 
 $T > T_{c}$ that condenses into the superfluid condensate when $T < T_{c}$.} 
\label{fig2} \end{figure}

\begin{figure}
\caption{The ratio of the spectral weight difference $[N_{N}(\omega) - 
N_{S}(\omega)]$ to the superfluid density $\rho_{s}$ as described in the 
text for optimally ($T_{c} = 81$~K) and over-doped ($T_{c} = 32$~K) Tl-
2201.  The superfluid condensate is accumulated from lower energies in the 
over-doped sample. To calculate the over-doped curve, we extrapolated 
(shown as a dashed line) $\sigma_{1}(\omega, T = 35$~K) from $\omega = 16$~cm$^{-1}$ to 
$\omega = 0$ using a 
Drude function with $\sigma_{dc} = 15$~$(\Omega$~cm$)^{-1}$ and 
$\tau^{-1} = 45 $~s$^{-1}$ and $\sigma_{1}(\omega,T = 5 K) = 5$~$ 
(\Omega $~cm$)^{-1}$.} \label{fig3} \end{figure} 


\begin{references}
\bibitem{footnote2} Data shown in Fig.~3 suggested a stronger suppression of 
the energy scale from which the condensate is collected than a factor of 3 
inferred from Eq.~3. This equation implies that the entire spectral weight 
contained in the normal state conductivity at $\omega<\Omega_S$ is 
transferred to the $\delta$-function in superconducting state. This is a 
correct assumption for a BCS superconductor, but in cuprates only a small 
fraction of the normal state spectral weight goes into condensate (see 
Fig.~2).
\end{references}
\end{document}